# Equilibrium tuned by a magnetic field in phase separated manganite

**M.Quintero[1,2] , F. Parisi[1,2] , G. Leyva[1,2] , and L. Ghivelder[3]**

[1] Departamento de Física, Comisión Nacional de Energía Atómica, Av. Gral Paz

1499 (1650) San Martín, Buenos Aires, Argentina

[2] Escuela de Ciencia y Tecnología, UNSAM, Alem 3901, San Martín, Buenos Aires, Argentina

[3] Instituto de Fisica, Universidade Federal do Rio de Janeiro, C.P. 68528, Rio de Janeiro, RJ

21941-972, Brazil.

Corresponding author: mquinter@cnea.gov.ar

We present magnetic and transport measurements on $La_{5/8-y}Pr_yCa_{3/8}MnO_3$ with $y$ = 0.3 , a manganite compound exhibiting intrinsic multiphase coexistence of sub-micrometric ferromagnetic and antiferromagnetic charge ordered regions. Time relaxation effects between 60 and 120K, and the obtained magnetic and resistive viscosities, unveils the dynamic nature of the phase separated state. An experimental procedure based on the derivative of the time relaxation after the application and removal of a magnetic field enables the determination of the otherwise unreachable equilibrium state of the phase separated system. With this procedure the equilibrium phase fraction for zero field as a function of temperature is obtained. The presented results allow a correlation between the distance of the system to the equilibrium state and its relaxation behavior.

## 1. Introduction

In earlier studies, the phase separated state in manganites was proposed to explain the colossal magnetoresistance found in this class of materials [1]. A lot of work was dedicated to show the existence of this state by different techniques [2,3,4,5] and, nowadays, this intrinsic tendency towards segregation is among the most important issues in the physics of strongly correlated systems, including others oxides with perovskite like structure, such as cobaltites and cuprates.[6] Although the origin of phase separation (PS) is not fully understood, both theoretical [7, 8, 9] and experimental [10] evidence point to the intrinsic disorder as main factor responsible for the stabilization of this inhomogeneous state. Due to the competition between the coexisting phases, some interesting time dependent effects have been observed, such as cooling rate dependence [11], relaxation [12,13,14], giant 1/f noise [15], and two-level fluctuations [16]. In addition, it is relatively easy to unbalance the relative phase fractions either via external stimuli [17,18,19], by inducing strains [14,20,21], or producing geometrical confinement[22].

The extremely rich variety of physical phenomena found in manganites has the systems $La_{1-x}Ca_xMnO_3$ and $Pr_{1-x}Ca_xMnO_3$ in permanent focus. In the former, for Ca doping level $x = 0.375$ the ferromagnetic (FM) ordering temperature $T_C$ is maximized [23]. Doping the optimized $La_{0.625}Ca_{0.375}MnO_3$ system with the smaller Pr ions introduces distortions and accommodation strains, which in turn determine the strong tendency towards electronic phase segregation [24]. The end members of the mixture $La_{0.625-y}Pr_yCa_{0.375}MnO_3$ exhibit nearly homogeneous ground states, FM for $y = 0$ and antiferromagnetic - charge ordered (AFM/CO) for $y = 0.625$. As a function of the Pr content $y$, the crossover between the FM and the AFM/CO ground states occurs around $y \approx 0.5$. In the range $0.3 < y < 0.4$ phase separation features fully develop. The compound with $y = 0.4$ deserved a deep and thorough study [7,20,25,26], mainly focused on the dynamic of the phase separated state. As previously shown, the slow growing evolution of the stable low temperature FM phase against the AFM/CO one is associated with a distribution of energy barriers, which in turns leads to the appearance of blocked metastable states. The presence of these blocked states are the responsible for the occurrence of abrupt field-induced jumps in the magnetization at very low temperatures [27,28,29] one of the most puzzling phenomenon in the study of metamagnetic transitions.

In many aspects, the dynamic of the phase separated state resembles that of the spin glass[30]. Not solely the existence of a slow dynamic, but also typical phenomena of spin glass as aging and rejuvenation have been observed [31, 32] But while spin glasses can relax indefinitely, the existence of a true phase separated equilibrium state in manganites establish an end point to the relaxation processes, regardless of the fact that it can or not be reached in laboratory times. The sign of the velocity of the time relaxation determines on what side of the equilibrium line the system is. For instance, if at a given temperature $T$ and magnetic field $H_0$ the system is in a state where the FM phase fraction is below its ($T$ and $H_0$ dependent) equilibrium value, there is an excess of the AFM/CO phase evolving to equilibrium. This process is characterized by a decrease of the resistivity and an increase of the magnetization as a function of time. On the other hand, if the actual FM phase fraction is above its equilibrium value, the excess of the FM phase will eventually transforms in to the AFM/CO phase. In the former case the application of a magnetic field $H > H_0$ will promote the increase of the FM phase fraction, leading eventually the system above the equilibrium line corresponding to the field $H_0$, with the consequent change of the sign of the velocity of relaxation once the magnetic field is returned back to its base value $H_0$. Therefore, there must be a threshold field $H_{th}$ which drives the sample to its equilibrium point for a particular $T$ and $H_0$ values. In other words, the field can be used to tune the system to its (otherwise unreachable) equilibrium state. It is worth noticing that a description of the phase separated state as composed by two different phases is oversimplified. It has been observed [20] that the non-FM region could be formed by more than one insulating phase. Intrinsic differences between the insulating phases, paramagnetic and AFM-CO , would be reflected in resistivity data, but hardly in magnetization measurements at moderated magnetic fields. For the sake of simplicity, we will refer to the phase separated state as formed by two phases, FM and non-FM, unless a more detailed description is necessary.

In the present study we use relaxation measurements tuned by a magnetic field to investigate the equilibrium state as a function of temperature of the phase separated manganite $La_{5/8-y}Pr_yCa_{3/8}MnO_3$, with $y$ = 0.3. We study a polycrystalline sample of this compound through resistivity and magnetization measurements. Metastable and out of equilibrium features were probed through time relaxation, both in transport and magnetic measurements. We show that, in a specific temperature range, equilibrium can be reached, irrespective of the slow dynamic of the system, through the application of an appropriate magnetic field. The determination of the magnetic field $H_{th}$ needed to bring the system to the steady state at a given field $H_0$ is the basis of the tuning process proposed for the identification of the equilibrium state.

## 2.Experimental details

High quality polycrystalline samples of $La_{0.325}Pr_{0.300}Ca_{0.375}MnO_3$ were synthesized by the sol-gel technique. Thermal treatments were performed at 1400 °C during 16 hours. Average grain size is around 2 microns, as observed from scanning electron microscopy. Magnetization measurements were performed in a Quantum Design PPMS system, as function of temperature, magnetic field, and time. Electrical transport measurements were performed by the standard four probe method in a home made system, with a closed cycle cryogenerator.

## 3.Results and discussion

Figures 1a and 1b characterize the behavior of the system, by displaying the resistivity ρ (*H*=0) and low field magnetization M (*H* = 6 mT) as a function of temperature, both on cooling and warming runs. Three regimes can be identified at different temperature ranges. In the high temperature region, $T > 220$ K, the system is in a paramagnetic - insulating state. In the intermediate temperature range, $60 < T < 120$ K, a non-fully FM state develops, signaling the presence of phase separation in this range of temperatures. At low temperatures, $T < 60$ K, the magnetization has a new sudden increase, leading to a low temperature state characterized by a plateau in magnetization, and a metallic behavior in resistivity.

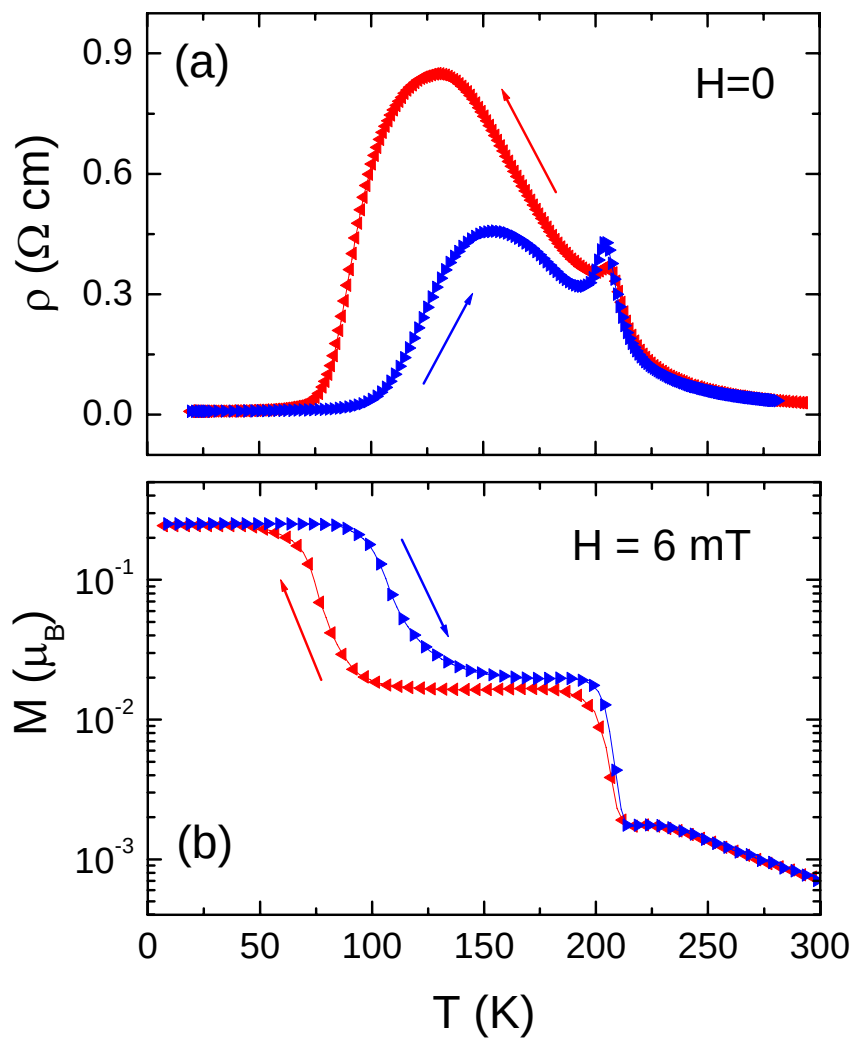

Figure 1: (a) resistivity and (b) magnetization of $La_{0.325}Pr_{0.300}Ca_{0.375}MnO_3$ on cooling and warming modes.

We have investigated the dynamic behavior of the system in the phase separated regime. Measurements were performed after field-cooled cooling from room temperature to different target temperatures in independent runs. The applied magnetic field was kept as low as possible, H=0 for resistivity, and H=0.006 T for magnetization experiments. . For temperatures below $T \approx$ 220 K time relaxation effects are present both in M and ρ. The logarithmic time dependence of the relaxation indicates a complex collective evolution of the system towards equilibrium. Figure 2a shows the normalized resistivity measured as a function of time for selected temperatures. Analogously, Fig. 2b shows the time dependence of the normalized magnetization. The temperature dependent transport and magnetic relaxation rates, also called resistive and magnetization viscosities, and defined as $S_R = \frac{1}{R_0}\left(\frac{dR}{d\log t}\right)$ and $S_M = \frac{1}{M_0}\left(\frac{dM}{d\log t}\right)$ respectively, where $R_0$ is the initial resistance and $M_0$ the initial magnetization, are displayed in Fig. 2c. The derivatives are taken at the highest measured *t* range. In this figure two regimes are clearly distinguished. Above 120 K, the negative values of $S_M$ and the positive values of $S_R$, indicate an excess of FM phase with respect to equilibrium.

In this regime it is worth noting the large values obtained of $S_R$ compared with that of $S_M$. This fact could be indicating the existence of relaxations not just from the FM to the non-FM phase but also internal relaxations within the insulating non-FM regions (for instance, from the paramagnetic to the CO-AFM phase) which are probed by resistivity measurements but not revealed by magnetization. On the other hand, below 120 K, the positive values of $S_M$ and the negative values of $S_R$ indicate an opposite trend, with a FM fraction lower than the the equilibrium one. In this temperature range the growth of the FM phase can be assisted by the application of an external magnetic field. This accelerates the relaxation process and opens a possibility to reach equilibrium in measurable times. We have used this fact to characterize the equilibrium state within the temperature range 60 < $T$< 120K. The procedure employed was as follows: we cooled the sample from room temperature to the target temperature under a field $H_0$=0. Once the temperature is stabilized the resistance is measured for 10 minutes, and the resistive viscosity $S_R(H=0)$ is determined. After that, a magnetic field $H_1 > H_0$ is applied for 5 minutes and subsequently removed. The new relaxation of the resistivity is measured and $S_R(H_1)$ is obtained, where $S_R(H_1)$ means "the viscosity measured at field $H_0$.after the application and removal of $H_1$". By repeating this procedure for different $H$ values we have obtained the data displayed in Fig. 3, for a target temperature of 110 K. It is clearly observed in this figure that the application of a threshold field $H_{th}$ = 0.23T has to drive the system close to the equilibrium at this temperature. In other words, the application and subsequent removal of a field $H_{th}$ promotes the growth of the FM phase to its zero field equilibrium value. To determine $H_{th}$ we have used resistivity measurements instead of magnetization, in order to determine the zero field equilibrium behavior of the sample.

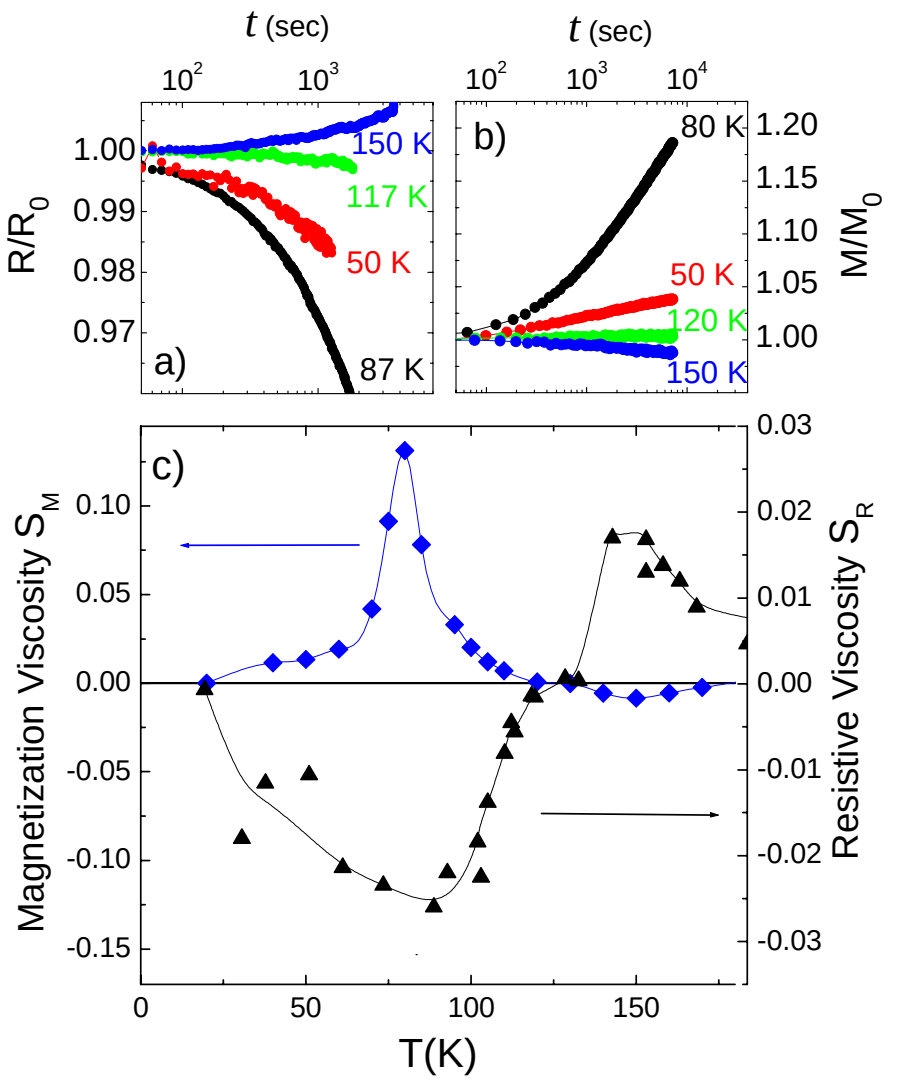

Figure 2: (a) Normalized resistivity (H = 0) and (b) magnetization (H = 6mT), measured as a function of time at selected temperatures; (c) relaxation rates: magnetization viscosity $S_M$ (∧) and resistive viscosity $S_R$ (7), defined as $S_R = \frac{1}{R_0}\left(\frac{dR}{d\log t}\right)$ and $S_M = \frac{1}{M_0}\left(\frac{dM}{d\log t}\right)$ respectively, where $R_0$ is the initial resistance and $M_0$ the initial magnetization, plotted as function of temperature. Each data point was measured on independent runs, cooling the sample from room temperature to the target temperature.

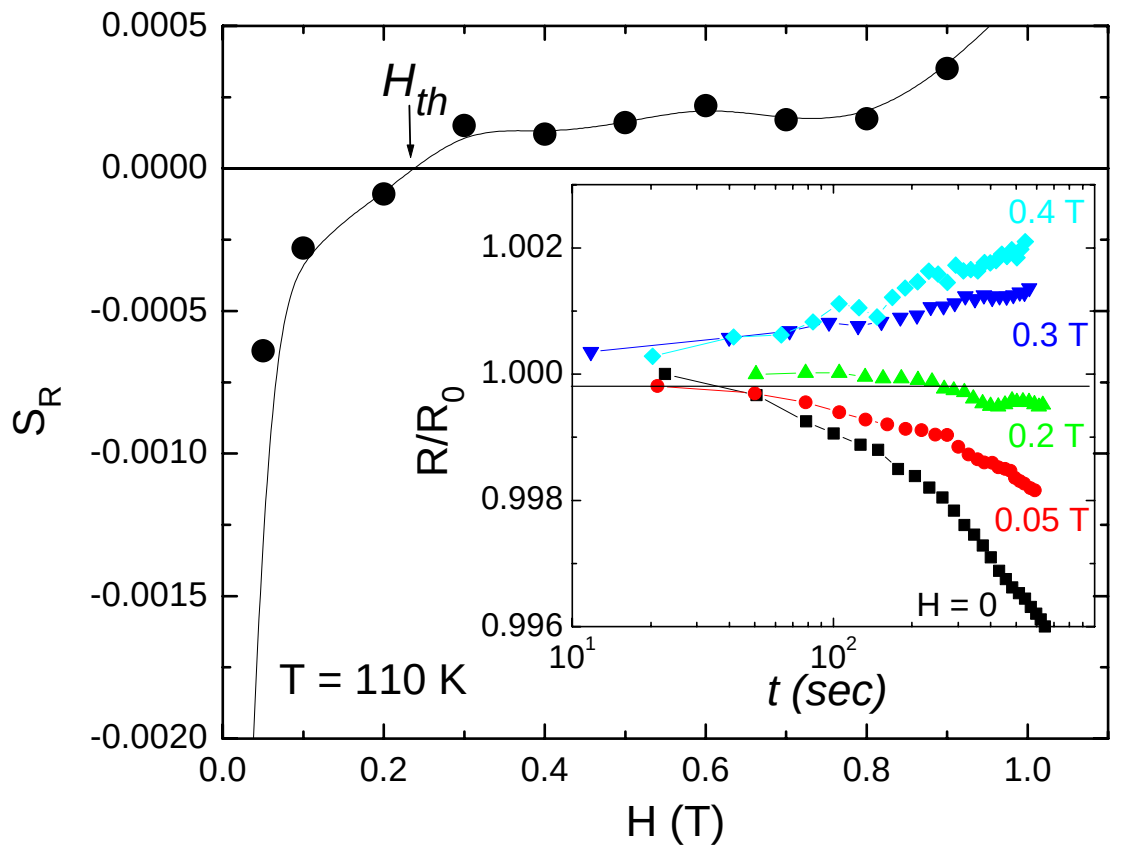

Figure 3: Resistive viscosity $S_R$, measured at $T$ = 110 K, after the application and removal of different magnetic fields. The inset shows the time evolution of the normalized resistance, after the application and removal of the applied fields. This data was used to calculate the values of $S_R$ plotted in the main panel. Measurements were performed after cooling the sample from room temperature to the target temperature at 2 K/min. (see text for details).

With this procedure it is possible to determine the equilibrium FM fraction $x_{eq}$(FM) at the given temperature from magnetization measurements as a function of field, M($H$). Figure 4 shows the M($H$) curve at 110 K. In the upward curve, the point marked with a star corresponds to the magnetization of the sample at $H_{th}$ after ZFC to $T$=110K. The star in the downward curve indicates the magnetization of the saturated FM sample at the same field. The quotient between both values of magnetization corresponds to the FM fraction obtained after application of $H_{th}$. As explained above, it has to be close to the zero field equilibrium FM fraction at 110K.

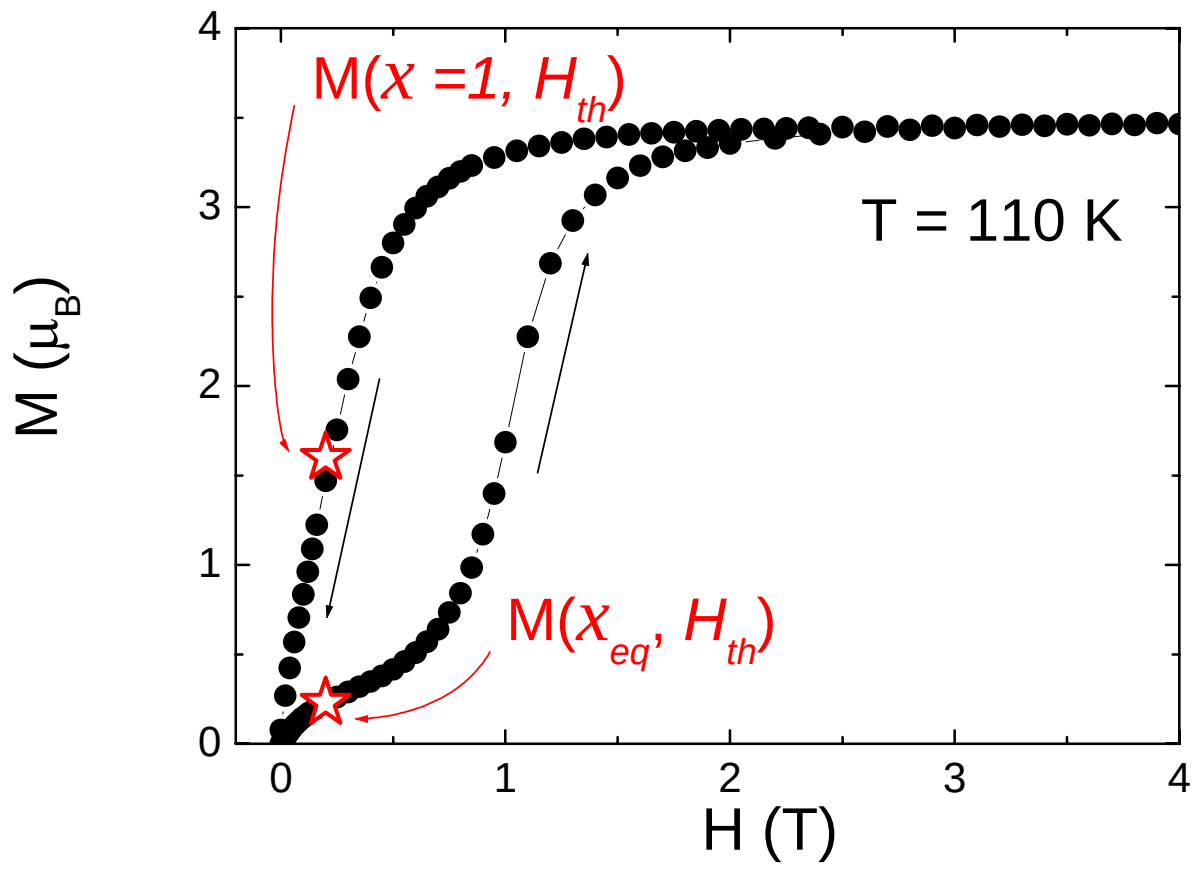

Figure 4: Magnetization as a function of applied magnetic field, M vs. $H$, measured at $T$ = 110 K. The stars indicate de values of M measured with a field $H_{th}$ on the upward and downward curves.

The above described procedure yield the determination of the equilibrium FM fraction as a function of temperatute, $x_{eq}$(T), plotted in Fig. 5a. For comparison, the experimental FM fraction, $x_{exp}$(T), is also plotted. The latter is obtained by measuring the M($T$) curve under a low field (6 mT). In the lower panel we display the temperature dependence of the field $H_{th}$ needed to drive the sample to the (zero field) equilibrium state, after ZFC to the target temperature. The equilibrium fraction $x_{eq}$(FM) goes from zero to 1 in a narrow temperature range, in which a true equilibrium phase separated state occur.

A description of the above presented results requires a qualitative understanding of the characteristic dynamic of the phase separates state, which can be phenomenologicaly described through a simple model of evolution through energy barriers with diverging height as the system approaches equilibrium. Although essentially qualitative in nature, this model was able to predict the dynamic $H$-$T$ phase diagram for a phase separated manganite. [25, 26] Following Ref. 25, the growth velocity of the FM phase against the AF/CO, assuming $x < x_{eq}$, is given by

$$\frac{dx}{dt} = \nu_0 e^{-\frac{U(H)}{(x_{eq}-x)kT}} \qquad (1)$$

where $\nu_0$ is a typical velocity of relaxation, $U(H)$ is the field-dependent barrier height, $x_0$ the FM phase fraction at the beginning of the relaxation, and $k$ is the Boltzman constant. Taking as variable $u = x - x_0$, and assuming that, in the early stages of the relaxation process, $u << x_{eq} - x_0$, we can integrate approximately the Eq. 1, leading to $x \approx x_0 + S\ln(t/t_0 + 1)$, with $S = \frac{(x_{eq} - x_0)^2 kT}{U(H)}$. Figure 5c displays the function $(x_{eq} - x_0)^2 T$ obtained from the data of Fig. 5a, and the magnetization viscosity from Fig 4. It can be seen a qualitative agreement between the measured and calculated behavior. Although it might not be enough to ensure the true function form for the velocity of evolution given in Eq. 1, it establishes a clear relation between the viscosity and the state of the system, i.e., the initial state given by $x_0$, the equilibrium $x_{eq}$, and temperature *T*. The data presented in Fig. 5b indicates that, due mainly to the enhancement of the blocking of metastable states, as the temperature is lowered larger magnetic fields are needed to lead the system to equilibrium. The linear relation observed between $H_{th}$ and *T* seems to indicate a linear relation between U(*H*) and *H*, but this statement must be corroborated by further experiments.

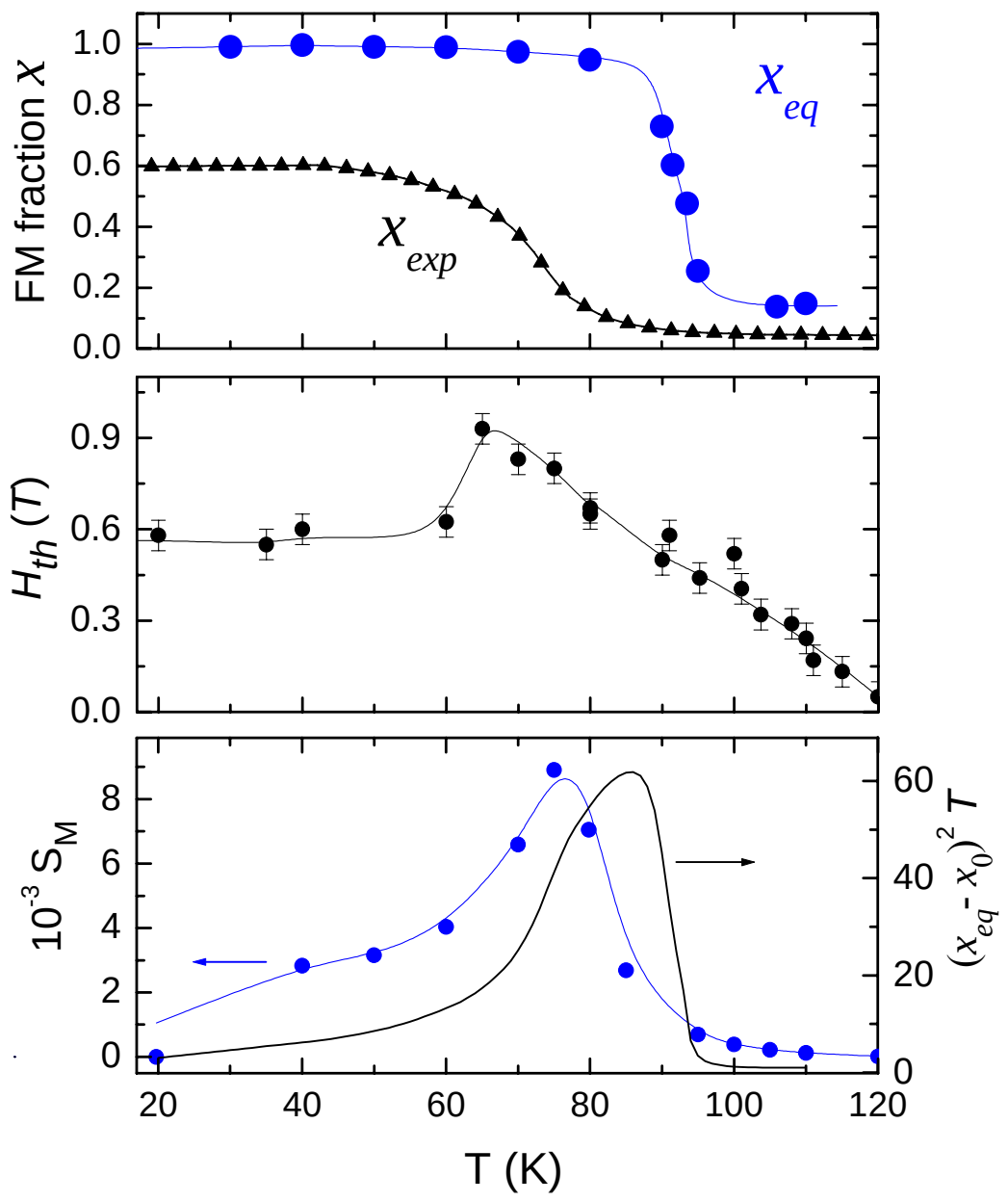

Figure 5: (a) equilibrium ($x_{eq}$) and experimental (H = 6 mT) ($x_{exp}$) ferromagnetic phase fractions as function of temperature; (b) threshold field $H_{th}$ required to obtain the equilibrium FM fraction; (c) magnetization viscosity compared with the estimated from the function $(x_{eq} - x_0)^2 T$.

## 4. Conclusions

In this investigation we have performed magnetic and transport measurements on $La_{0.325}Pr_{0.300}Ca_{0.375}MnO_3$ manganite compound, focusing our attention in the features related with the dynamics of the phase separation between 60 and 120K. A systematic study of the time relaxation properties was performed and discussed. Magnetization and resistivity viscosities were obtained. The slow logarithmic relaxation observed were accounted for using a collective dynamic model with diverging energy barriers as the system approaches equilibrium. We have developed an experimental procedure to obtain the equilibrium state of the system in the phase separated regime, which allows us to determine the zero field equilibrium FM fraction in a wide temperature range. The correlation between the distance of the system to the equilibrium state in a conventional field cooled cooling experiement and the relaxation behavior was established. Our results indicates that the knowledge of the true equilibrium state is a *sine qua non* condition to further investigate, experimental or theoretically, the properties of phase separated state.

## Acknowledgements

M. Quintero is also member of CIC, CONICET. This work was supported by the project CAPES-Secyt 121/07, with. additional support from FAPERJ, and CNPq (Brazil). Helpful discussions with P. Levy are gratefully acknowledged.